\documentclass[
    ,final            
  ]
  {aipproc}

\layoutstyle{6x9}

\begin{document}

\title{Giant Steps in Cefal\`u}

\classification{05.10.Ln, 42.68.Ay, 95.30.Jx, 97.10.Ex, 97.60.Bw}

\keywords      {Monte Carlo methods, radiative transfer, supernovae}

\author{David J.~Jeffery}{
  address={Homer L. Dodge Department of Physics \& Astronomy, University of Oklahoma,
                440 W. Brooks St., Norman, Oklahoma 73019, U.S.A.},
}

\author{Paolo A.~Mazzali}{
  address={INAF-Osservatorio Astronomico di Trieste, Via G.B.~Tiepolo 11, 34131 Trieste, Italy},
}



\begin{abstract}
   {\it Giant steps} is a technique to accelerate Monte Carlo radiative transfer
   in optically-thick cells (which are isotropic and homogeneous in matter properties and 
   into which astrophysical atmospheres are divided)
   by greatly reducing the number of Monte Carlo steps needed to propagate photon packets through such cells.
   In an optically-thick cell, packets starting from any point (which can be regarded a point source)
   well away from the cell wall act essentially as packets diffusing from the point source in 
   an infinite, isotropic, homogeneous atmosphere. 
   One can replace many ordinary Monte Carlo steps that a packet diffusing from the point source takes by a randomly directed 
   giant step whose length is slightly less than the distance to the nearest cell wall point from the point source.
   The giant step is assigned a time duration equal to the time for the RMS radius for a burst of packets diffusing from the point source 
   to have reached the giant step length. 
   We call assigning giant-step time durations this way RMS-radius (RMSR) synchronization.
   Propagating packets by series of giant steps in giant-steps random walks in the interiors of optically-thick cells constitutes 
   the technique of {\it giant steps}.   
   {\it Giant steps} effectively replaces the exact diffusion treatment 
   of ordinary Monte Carlo radiative transfer in optically-thick cells by an approximate diffusion treatment.
   In this paper, we describe the basic idea of {\it giant steps} and report
   demonstration giant-steps flux calculations for the grey atmosphere. 
   Speed-up factors of order 100 are obtained relative to ordinary Monte Carlo radiative transfer.
   In practical applications, speed-up factors of order ten \cite{mazzali2001} and perhaps more are possible.
   The speed-up factor is likely to be significantly application-dependent and there is a trade-off between speed-up and accuracy.
   This paper and past work \cite{mazzali2001} suggest that 
   giant-steps error can probably be kept to a few percent 
   by using sufficiently large boundary-layer optical depths
   while still maintaining large speed-up factors.
   Thus, {\it giant steps} can be characterized as a moderate accuracy radiative transfer technique. 
   For many applications, the loss of some accuracy may be a tolerable price to pay for 
   the speed-ups gained by using {\it giant steps}.
\end{abstract}

\maketitle



\section{INTRODUCTION}

Monte Carlo radiative transfer simulates radiation transfer by propagating photon
packets (which represent aggregations of photons) through model atmospheres.
The atmospheres are usually divided into cells that are isotropic and homogeneous in
matter properties:   e.g., slab cells that are infinite in 2 dimensions for plane-parallel
atmospheres or concentric spherical shells for spherically symmetric atmospheres. 
From the packet-matter interactions, one can calculate cell thermal states
and from the packets that escape the atmosphere, the emergent flux and spectrum.   
Since there are almost always vastly fewer packets than actual photons, one is doing
a statistical sampling to determine results and these results will have errors 
due to statistical fluctuations. 
The statistical errors can be reduced by increasing the number of packets. 

    Monte Carlo radiative transfer has powerful features. 
Packet-matter interactions and complex atmospheres in multi-dimensions can usually be treated straightforwardly 
with great physical realism.
Thus, one can avoid many of the approximations or complex treatments needed for other radiative
transfer techniques.
Monte Carlo radiative transfer codes are often easy to develop in comparison to other radiative transfer codes.
Monte Carlo radiative transfer is embarrassingly parallelizable in that one can propagate
as many packets through an atmosphere simultaneously as one has processors:  the effects
of the packets merely have to be summed at the end of the propagation calculation.
Monte Carlo radiative transfer gives robust convergence in self-consistent atmosphere
calculations where an atmosphere solution is iterated to convergence via alternating 
radiative transfer and thermal state calculations \cite{lucy1999a,kasen2006}.  

    Monte Carlo radiative transfer also has severe disadvantages relative to other radiative
transfer techniques.
In doing Monte Carlo radiative transfer, it can be CPU time expensive to reduce the statistical errors to an acceptable level. 
Another (but related) problem is that Monte Carlo radiative transfer can be
very slow for optically-thick atmospheres which
in the deep interior will usually be divided into optically-thick cells.
The slowness is caused, of course, by having to propagate each packet 
through all its many interactions as it traverses the optically-thick cells.
The details of a packet's propagation are not of interest, but 
in the ordinary Monte Carlo treatment one must calculate them nevertheless. 

    Can anything be done about the slow propagation in optically-thick cells?
Yes.
Take giant steps.   


\section{THE BASIC IDEA OF GIANT STEPS}

While mulling on Monte Carlo radiative transfer on the night
of 2006 February~13, one of us (D.J.J.) had a eureka moment.  
Consider a packet (he said to himself) in an optically-thick cell of a pure-isotropic-scattering atmosphere 
just after a scattering event.
The packet will random walk a long way in the cell and will always be in a random direction relative to
the start point:  
as long as the packet does not leave the cell, its motion is just as if it were in an
infinite, isotropic, homogeneous atmosphere.
Why not just ``giant-step'' the packet from the start point in a random direction as far as one can go in the
cell consistent with the packet behaving as if in an infinite, isotropic, homogeneous atmosphere.
For the consistency, the giant-step length should not exceed (and perhaps be a bit less (see below)) than the 
distance to the nearest cell wall point:  but note the giant step is in a random direction, not to the nearest cell wall point.
After one giant step, take another giant step in a random direction, and so on in a giant-steps random walk.
With this random walk, the packet is propagated a long way quickly inside the cell:  there should be a
great speed-up relative to ordinary Monte Carlo propagation.


\begin{figure}
  \includegraphics[height=.5\textheight]{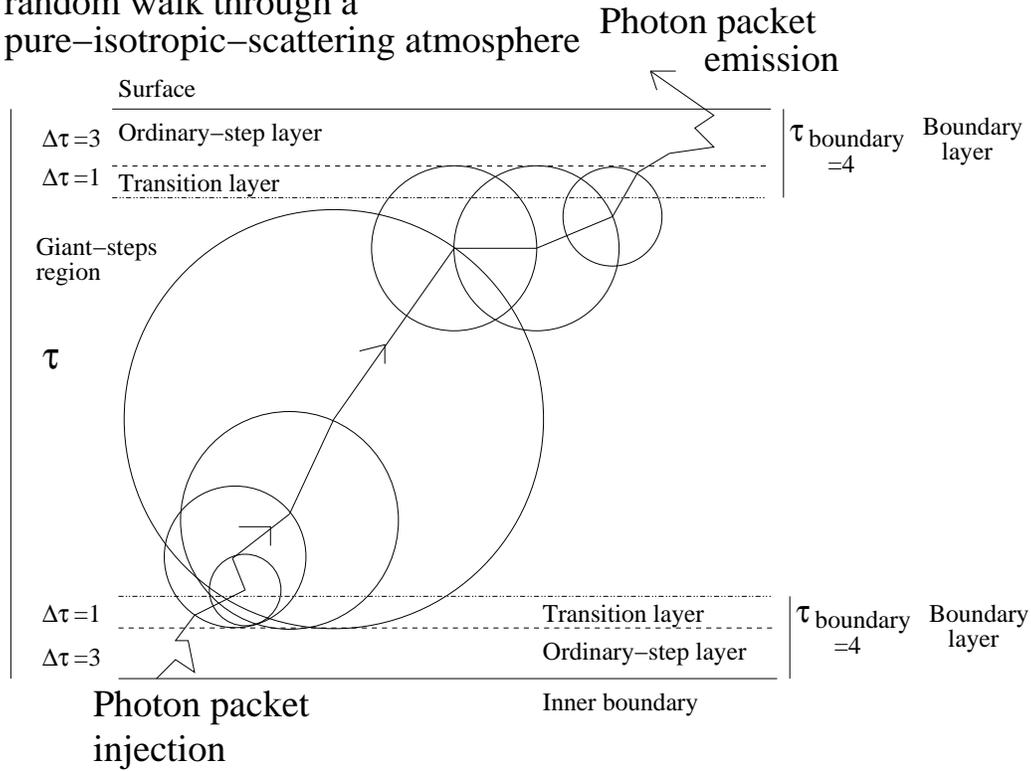}
  \caption{A cartoon of a giant-steps random walk in a plane-parallel, pure-isotropic-scattering atmosphere consisting of one cell
   of optical depth $\tau$.
   The circles represent the spheres that are the loci formed from the giant-step endpoints possible 
   for each start point in the giant-steps region.
   A giant step is taken in a random direction to the sphere from a start point.}
\label{fig-giant-steps-random-walk}
\end{figure}

     Of course, how does the packet get out of the cell and if the packet is less than optical depth~1 from the
cell wall, the giant steps turn into baby steps. 
Ah, change to ordinary Monte Carlo propagation when one is close to the cell wall:  ordinary Monte Carlo
steps have variable length determined by a cumulative probability distribution and a random direction.
The ordinary Monte Carlo steps allow the packet to leave the cell or re-enter the giant-steps region (the interior
region of the cell where only giant steps can be started.) 
It turns out that one needs a boundary layer (of optical depth $\tau_{\rm boundary}$)
consisting of an ordinary-step layer 
(where only ordinary steps are taken) 
of optical depth a few adjacent to the cell wall and transition layer of optical depth 1 adjacent to the ordinary-step layer.
Only ordinary steps are started in the boundary layer, but giant steps penetrate into the transition layer from the 
giant-steps region.
With the boundary-layer setup, the giant-step length is determined by the distance to the nearest ordinary-step layer point, not
as in first conception to the nearest cell wall point.
Since the transition layer has optical depth 1, there are no baby steps.
In fact, the ordinary-step layer can be shrunk to zero width, but for high accuracy it should have an optical depth of a few
(see the next section).
Figure~\ref{fig-giant-steps-random-walk} shows
a cartoon of a giant-steps random walk through a plane-parallel, pure-isotropic-scattering atmosphere consisting of one cell.


     Whenever one has a smart idea there are really only two possibilities:  (a)~one is wrong;  (b)~someone has thought
of it before.
{\it Giant steps} (as we call the technique described above)
turned out to be case~(b) when another of us (P.A.M.) admitted to having used {\it giant steps}
(without calling it that) surreptitiously in supernova lightcurve calculations \cite{mazzali2001}.
A full presentation of {\it giant steps} with discussion of how it applies to general atmospheres including, of course,
multi-frequency treatment and energy deposition (which is needed for self-consistent atmosphere solutions)
is given by \cite{jeffery2007}.
Here we will elaborate just a bit on the description of {\it giant steps} given above.

    Some thought (aided by some calculations) shows that {\it giant steps} is approximating
packet diffusion in a cell 
rather than treating that diffusion exactly as in ordinary Monte Carlo radiative transfer.
Consider two isotropic-scattering packets diffusing in an infinite, isotropic, homogeneous atmosphere.
One packet starts diffusing from a point 
and the other packet starts diffusing from giant-step length away in a random direction
starting an appropriate time duration later.
The packets will have approximately the same average behavior.
This can be seen by imagining each packet as part of a isotropic packet burst:  
the first from a point source;  
the second from a sphere source of radius the giant-step length. 
The behaviors of the bursts converge with time.
This convergence is optimum if the time duration assigned to the giant step is the time for the
root-mean-square (RMS) radius of the point-source burst to reach the giant-step length.
We call setting the giant-step time durations this way RMS-radius (RMSR) synchronization.
RMSR synchronization is the optimum synchronization for {\it giant steps} \cite{jeffery2007}.
A key aspect of RMSR synchronization is that the RMS radii of point- and sphere-source bursts
are always equal after the giant-step time duration.
The point- and sphere-source bursts 
give a heuristic picture that allows one to understand {\it giant steps}.
In actual calculations, one packet at a time (at least on one processor) is propagated through
its whole trajectory in the atmosphere and the packets take series of giant steps in optically-thick cells.

\section{DEMONSTRATION GIANT-STEPS FLUX CALCULATIONS}

\begin{figure}
  \includegraphics[height=.6\textheight,angle=-90]{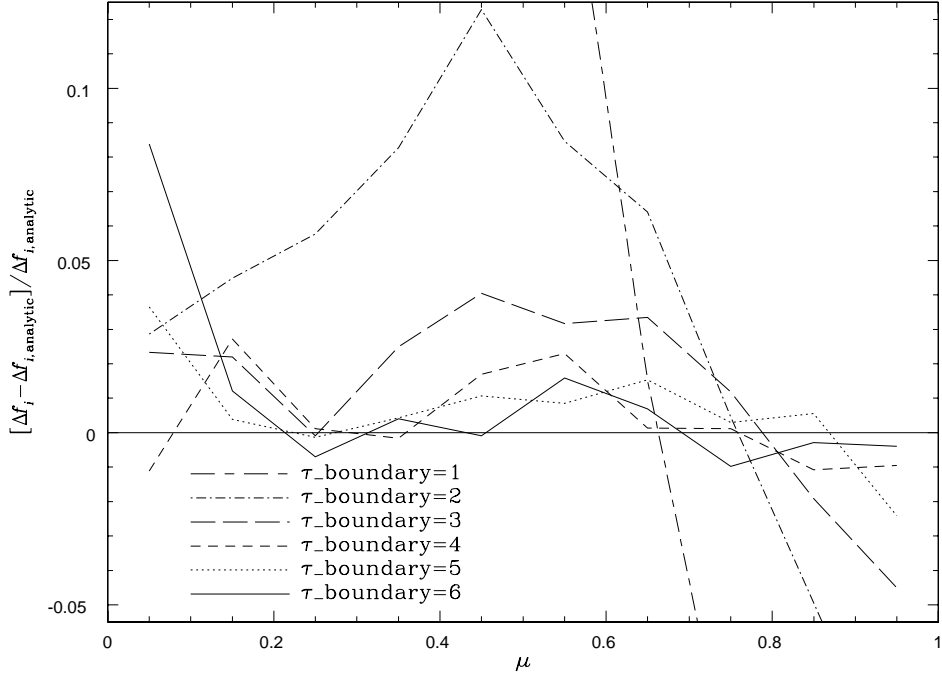}
  \caption{Relative deviations for $\Delta f_{i}$ indicated at $\mu$-interval midpoints by connecting lines.}
\label{fig-relative-deviation}
\end{figure}

Counting giant-step time durations is essential in time-dependent giant-steps calculations and 
in any   
giant-steps calculations
where energy deposition needs to be done (i.e., in any self-consistent atmosphere solution calculations). 
However, for just propagating packets through the grey atmosphere 
(a time-independent, isotropic-scattering,
semi-infinite, plane-parallel atmosphere
with frequency-independent opacity and which can be treated as a pure-isotropic-scattering
atmosphere with effectively a single frequency:  \cite[e.g.,][p.~53--76]{mihalas1978})
in order to get the emergent flux, time
durations are not counted and one just does giant-step propagation of packets as described in the last section and as illustrated in
Fig.~\ref{fig-giant-steps-random-walk}. 
We have done such giant-steps flux calculations for the grey atmosphere with optical depth 1000 (and inner boundary
treated in the diffusion approximation) treated as one cell and with $10^{8}$ packets injected at the inner boundary. 
From the exact analytic solution of the grey atmosphere \cite[e.g.,][p.~71--73]{mihalas1978}, 
we know that only about $1.3\times10^{5}$ packets will escape from the surface
of the atmosphere:  the others are absorbed at the inner boundary.  
As an example result, we consider the angular distribution of the emergent radiation field.
Because it is convenient to evaluate in a Monte Carlo calculation, we have chosen to represent 
the angular distribution by relative partial astrophysical fluxes given by
\begin{equation}
   \Delta f_{i}=2\int_{\mu_{i-1}}^{\mu_{i}} {I(\mu)\over F}\mu\,d\mu \,\, ,
\end{equation}
where $\mu$ is the direction cosine for the angle from the normal to the surface,
$i$ is an index (running $1,2,\ldots$),
$\mu_{i}-\mu_{i-1}$ is the $\mu$ interval for which $\Delta f_{i}$ is evaluated, and $I(\mu)/F$ is the 
specific intensity divided by the astrophysical flux (i.e., the limb-darkening law).   
We evaluated $\Delta f_{i}$ for ten $\mu$ intervals of $0.1$ spanning the full range for emergent flux
from $\mu=0$ to $\mu=1$
for a set of giant-steps calculations with varied boundary-layer optical depth $\tau_{\rm boundary}$. 
To test for accuracy, we evaluated the relative deviations of the giant-steps $\Delta f_{i}$ values from ones
calculated from the exact analytic limb-darkening law for the grey atmosphere.
(The exact analytic limb-darkening law formula 
must 
be evaluated numerically:  tabulations are given by, e.g., \cite[p.~72]{mihalas1978} and \cite[p.~135]{chandrasekhar1960}.)
In Fig.~\ref{fig-relative-deviation},
we have indicated these relative deviations at the $\mu$-interval midpoints by 
connecting the values by lines:  this is clearer than plotting multiple sets of data by data points or in a histogram.

    From Fig.~\ref{fig-relative-deviation}, we see that the accuracy is poor for $\tau_{\rm boundary}\leq2$ for
which some relative deviations are off the plot. 
The maximum absolute value of a relative deviation is $1.29$ at 
$\mu=0.05$ for $\tau_{\rm boundary}=1$.
Note, however, that the net flux for $\tau_{\rm boundary}=1$ is in error by only $3\,$\% \cite{jeffery2007}:
thus, it is only the angular distribution that is rather inaccurate. 
The accuracy of the giant-steps calculations generally 
improves as $\tau_{\rm boundary}$ is increased up to about $\tau_{\rm boundary}=6$ where error is a few percent or less.
For higher values of $\tau_{\rm boundary}$, there is little significant improvement since the systematic error has
become smaller than the Monte Carlo statistical error.
The statistical error, in fact, increases as $i$ decreases  because the number of escaping packets 
per $\mu$ coordinate decreases as $\mu$ goes to $0$.
This explains the peak at 
$\mu=0.05$ for $\tau_{\rm boundary}=6$:  the peak is only a $2\sigma$ deviation, in fact.

    The giant-steps calculations are much faster than the counterpart ordinary Monte Carlo calculations.
The speed-up factors range from $\sim 250$ for $\tau_{\rm boundary}=1$ to $\sim 70$ for $\tau_{\rm boundary}=6$.
Obviously, the speed-up factor decreases as the boundary layer (where ordinary steps are taken) increases.

    Reader, conclusions are given in the abstract.
We have run out of room here. 



\begin{figure}
  \includegraphics[height=.5\textheight,angle=0]{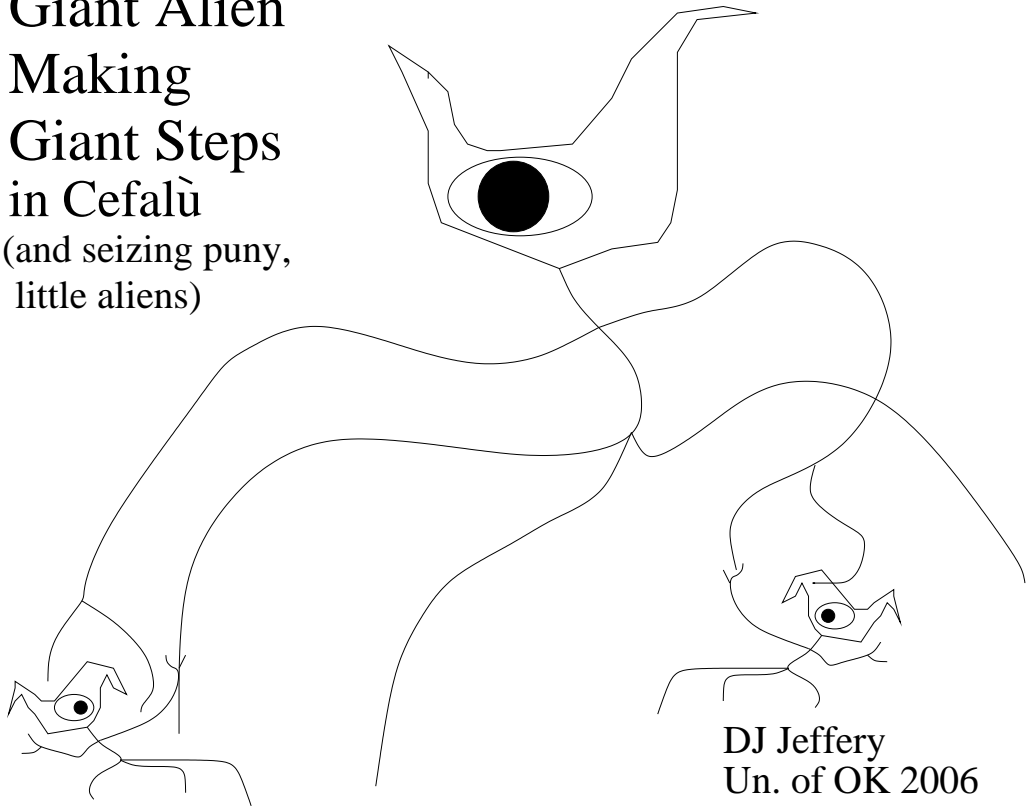}
  \caption{Giant Alien making giant steps in Cefal\`u.}
\label{fig-alien-giant}
\end{figure}


\begin{theacknowledgments}
The first and second authors independently came up with the idea of {\it giants steps}:  P.A.M. first and D.J.J. second.
Support for this work has been provided by NASA grant NAG5-3505 and
NSF grant AST-0506028.
We thank Eddie Baron and David Branch for comments and the conference organizers and staff for great days in Cefal\`u.

    Since Hale Bradt of the scientific organizing committee asked participants for a personal image 
in the talks and we had no images at all in Cefal\`u,  
we have included Fig.~\ref{fig-alien-giant} which illustrates the more fearful aspects of {\it giant steps}.

\end{theacknowledgments}



\bibliographystyle{aipproc}   

\end{document}